\documentclass[%
 aip,
 amsmath,amssymb,
 reprint,%
]{revtex4-1}

\usepackage{graphicx}% Include figure files
\usepackage{dcolumn}% Align table columns on decimal point
\usepackage{bm}% bold math
\usepackage[utf8]{inputenc}
\usepackage[T1]{fontenc}
\usepackage{mathptmx}
\usepackage{enumitem}
\usepackage{color}
 \usepackage[colorlinks=true,linkcolor=blue,citecolor=blue,urlcolor=blue]{hyperref}

\def\qe{\textsc{Quantum ESPRESSO}}
\def\MaX{\textsc{MaX}}

\usepackage[normalem]{ulem}
\usepackage[dvipsnames]{xcolor}
\definecolor{tangerine}{rgb}{0.944,0.522,0}
\definecolor{brown}{rgb}{0.633,0.156,0.156}

\newcommand{\editor}[2]{%
  \expandafter\newcommand\csname #1note\endcsname[1]{%
    \textcolor{#2}{(\textbf{#1:} ##1)}}%
  \expandafter\newcommand\csname #1\endcsname[1]{%
    \textcolor{#2}{##1}}%
  \expandafter\newcommand\csname #1cancel\endcsname[1]{%
    \textcolor{#2}{\sout{##1}}}%
  \expandafter\newcommand\csname #1change\endcsname[2]{%
    \textcolor{#2}{\sout{##1} ##2}}%
  \newenvironment{#1text}{\color{#2}}{\color{black}}
}

\editor{RV}{red}

\begin{document}% ****** Start of file aipsamp.tex ******

\title[\qe]{
  Quantum ESPRESSO toward the exascale}
\author{Paolo Giannozzi}
\affiliation{Dipartimento di Scienze Matematiche, Informatiche e Fisiche,
  Universit\`a di Udine, via delle Scienze 206, I-33100 Udine, Italy, European Union}
\affiliation{CNR-IOM, Istituto dell'Officina dei Materiali, SISSA, I-34136 Trieste, Italy, European Union}
\affiliation{Quantum ESPRESSO Foundation, Cambridge Road Ind Estate, Milton, Cambridge, CB24 6AZ, United Kingdom}

\author{Oscar Baseggio}
\affiliation{SISSA -- Scuola Internazionale Superiore di Studi Avanzati, via Bonomea 265, I-34136, Trieste, Italy, European Union}

\author{Pietro Bonfà}
\affiliation{Dipartimento di Scienze Matematiche, Fisiche e Informatiche, Universit\`a di Parma, Parco Area delle Scienze 7/A, I-43124 Parma, Italy, European Union}
\affiliation{Centro S3, CNR-Istituto Nanoscienze, via Campi 213/A, I-41125 Modena, Italy, European Union}

\author{Davide Brunato}
\affiliation{SISSA -- Scuola Internazionale Superiore di Studi Avanzati, via Bonomea 265, I-34136, Trieste, Italy, European Union}

\author{Roberto Car}
\affiliation{Department of Chemistry, Princeton University, Princeton, NJ 08544, USA}

\author{Ivan Carnimeo}
\affiliation{SISSA -- Scuola Internazionale Superiore di Studi Avanzati, via Bonomea 265, I-34136, Trieste, Italy, European Union}

\author{Carlo Cavazzoni}
\affiliation{CINECA - Via Magnanelli 6/3, I 40033 Casalecchio di Reno, Bologna, Italy, European Union}
\affiliation{Quantum ESPRESSO Foundation, Cambridge Road Ind Estate, Milton, Cambridge, CB24 6AZ, United Kingdom}

\author{Stefano de Gironcoli}
\affiliation{SISSA -- Scuola Internazionale Superiore di Studi Avanzati, via Bonomea 265, I-34136, Trieste, Italy, European Union}
\affiliation{CNR-IOM, Istituto dell'Officina dei Materiali, SISSA, I-34136 Trieste, Italy, European Union}

\author{Pietro Delugas}
\affiliation{SISSA -- Scuola Internazionale Superiore di Studi Avanzati, via Bonomea 265, I-34136, Trieste, Italy, European Union}
\affiliation{Quantum ESPRESSO Foundation, Cambridge Road Ind Estate, Milton, Cambridge, CB24 6AZ, United Kingdom}

\author{Fabrizio Ferrari Ruffino}
\affiliation{CNR-IOM, Istituto dell'Officina dei Materiali, SISSA, I-34136 Trieste, Italy, European Union}

\author{Andrea Ferretti}
\affiliation{Centro S3, CNR-Istituto Nanoscienze, via Campi 213/A, I-41125 Modena, Italy, European Union}

\author{Nicola Marzari}\affiliation{Theory and  Simulation  of  Materials (THEOS), and  National  Centre for  Computational  Design and Discovery of Novel Materials (MARVEL), \'Ecole Polytechnique F\'ed\'erale de Lausanne (EPFL), CH-1015 Lausanne, Switzerland}
\affiliation{Quantum ESPRESSO Foundation, Cambridge Road Ind Estate, Milton, Cambridge, CB24 6AZ, United Kingdom}

\author{Iurii Timrov}
\affiliation{Theory and  Simulation  of  Materials (THEOS), and  National  Centre  for  Computational  Design and Discovery of Novel Materials (MARVEL), \'Ecole Polytechnique F\'ed\'erale de Lausanne (EPFL), CH-1015 Lausanne, Switzerland}

\author{Andrea Urru}
\affiliation{SISSA -- Scuola Internazionale Superiore di Studi Avanzati, via Bonomea 265, I-34136, Trieste, Italy, European Union}

\author{Stefano Baroni}
\affiliation{SISSA -- Scuola Internazionale Superiore di Studi Avanzati, via Bonomea 265, I-34136, Trieste, Italy, European Union}
\affiliation{CNR-IOM, Istituto dell'Officina dei Materiali, SISSA, I-34136 Trieste, Italy, European Union}
\affiliation{Quantum ESPRESSO Foundation, Cambridge Road Ind Estate, Milton, Cambridge, CB24 6AZ, United Kingdom}

\begin{abstract}
    \qe\ is an open-source distribution of computer codes for quantum-mechanical materials modeling, based on density-functional theory, pseudopotentials, and plane waves, and renowned for its performance on a wide range of hardware architectures, from laptops to massively parallel computers, as well as for the breadth of its applications. In this paper we present a motivation and brief review of the ongoing effort to port \qe\ onto heterogeneous architectures based on hardware accelerators, which will overcome the energy constraints that are currently hindering the way towards exascale computing.
\end{abstract}

\date{\today}

\maketitle

\section{Introduction}

% New heterogeneous computer architectures, based on multi-core chips enhanced with multiple hardware  ``accelerators'' are bringing about unprecedented  performances coupled with an acceptable energy consumption. 
The goal of this manuscript is to describe recent and ongoing work on the \qe{} software distribution for first-principle atomistic simulations. We focus in particular on the challenges posed by the new heterogeneous architectures, based on multi-core chips enhanced with multiple hardware ``accelerators'', coupling exceptional performances to an acceptable energy consumption. 
The large-scale adoption of these emerging architectures across different classes of computing systems will bring a paradigm shift into scientific computing, similar to what vector machines caused in the 80's, parallel machines in the 90's, massively parallel machines more recently. 

This paper is organized as follows.

In Sec.~\ref{sec:qe-at-the-turn}, we briefly describe the history and the current status of \qe{}. We give a short overview of its features and capabilities, by mostly referring to the relevant literature. 

In Sec.~\ref{sec:towards-the-exascale} we describe the challenges posed and opportunities offered by heterogeneous architectures. The opportunity is to reach what is dubbed ``exascale computing'': an unprecedented amount of computer power, opening new perspectives to computer simulations. The challenges, for scientific software in general and for \qe{} in particular, are unprecedented as well. The amount of needed changes is much larger, and the effects of changes much deeper, than in previous transitions to vector, parallel, massively parallel architectures. Moreover, several competing architectures are hitting the marketplace, each one coming with a different software stack and tools. 

Sec.~\ref{subsec:performance_portability} describes ongoing work towards {\em performance portability}, that is: the ability to obtain comparably high performances on different computer architectures, minimizing the need for maintaining hardware-specific code versions. The work of Sec.~\ref{subsec:performance_portability}, that focuses on low-level libraries, would however be futile without more work at a higher level, aimed towards future maintainability of a large and complex scientific software project. 

Sec.~\ref{subsec:SustainableSoftwareDevelopment} 
deals with recent and ongoing work aiming at a {\em sustainable development model}, that is: restructuring codes in a way that makes them easier to maintain, to extend, and especially to port to other architectures in a foreseeable future.

Sec.~\ref{subsec:GPUVersion} describes the current status, capabilities, and some benchmarks for \qe{} on NVIDIA graphics processing units (GPUs), one of the leading candidate architecture for the future ``exascale computer''. The benchmarks are designed for relatively small machines and do not aim at showing performances on large-sized systems. They aim instead at pointing out bottlenecks, inefficiencies, and the minimum size of calculations that saturate the computational power of a GPU.

Finally, Sec.~\ref{sec:Conclusions} contains our conclusions: a brief analysis of the achieved results and an outlook on forthcoming actions.

\section{\qe\ at the turn of the twenties}
\label{sec:qe-at-the-turn}

The \qe\ project was started in 2002, with the merger of three packages for density-functional theory (DFT) simulations using plane waves and (ultrasoft) pseudopotentials, which had been under development since the mid-80's:
\begin{itemize}[noitemsep,nolistsep]
\item[--] \texttt{PWscf}: code \texttt{pw.x} for self-consistent field (SCF) solution of Kohn-Sham equations and structural optimization, code \texttt{ph.x} for lattice-dynamical calculations using linear response, plus many other utility codes;
\item[--] \texttt{CP}: a code performing first-principle molecular dynamics (MD)
simulations of the Car-Parrinello type, specialized to large supercells;
\item[--] \texttt{FPMD}: a code similar to \texttt{CP}, but with a different and partially
overlapping  set of functionalities.
\end{itemize}

Innovation in quantum-mechanical materials modeling has always been one of the main concerns and distinctive features of this project. In particular, the linear-response code \texttt{ph.x} contained in \texttt{PWscf} was, to the best of the authors' knowledge, the first implementation~\cite{BGT} of density-functional perturbation theory (DFPT).~\cite{Giannozzi:1991, Baroni:2001,Gonze:1995,Gonze:1995b} The elimination of virtual orbitals from linear-response and excited-state calculations, pioneered in this implementation, has remained one of the distinctive features of \qe, later to be adopted by several other codes. On the same line, \texttt{CP} and \texttt{FPMD} were both offsprings of the first implementation\cite{CP} of Car-Parrinello MD, now merged into a single code, \texttt{cp.x}.

Since the beginning, \qe\ was released under an open-source license (GNU GPL) \footnote{At the time, \texttt{PWscf} had already been released under the GPL} and was conceived as a ``distribution'': an integrated suite of packages, following loose programming guidelines, rather than a monolithic application. The rationale for these choices lies in the philosophy of fostering collaboration and sharing of code among scientists, creating a community of users and developers, while pursuing innovation in methods  and algorithms. For more on such aspect, we refer to Refs. \onlinecite{QE1} and \onlinecite{QE2}.

% excerpt from  Nuovo Assaggiatore:
% writing software for use by other people requires a significant additional effort 
% with respect to writing software for one's own usage. This comes from the need to
% provide user and developer documentation and to guarantee the stability of the APIs.
% The scientific community should attribute more dignity to software
% development and give more reward to the people doing code development.

%\subsection{\qe\ today}

The \qe\ distribution has much evolved over the years. On the one hand,  more sophisticated theoretical methods and algorithmic advances have been implemented, in particular:
\begin{itemize}[noitemsep,nolistsep]
\item[--] projector-augmented waves (PAW);
\item[--] non-collinear and spin-orbit coupling (SOC) calculations;
\item[--] Hubbard-corrected functionals;
\item[--] nonlocal functionals and other semi-empirical and less-empirical
  corrections for van der Waals bonded systems;
\item[--] hybrid functionals, also exploiting orbital localization.
\end{itemize}

On the other hand, \qe\ has extended its original scope with additional
packages for more property calculations. We mention in particular:
\begin{itemize}[noitemsep,nolistsep]
\item[--] activation energies and reaction barriers using the
  nudged-elastic-band (NEB) method;
\item[--] superconducting transition temperatures, electrical resistivity, and other effects related to electron-phonon interactions;
\item[--] phonon linewidths, thermal conductivity, and anharmonic effects;
\item[--] nuclear magnetic resonance (NMR) chemical shifts and electron paramagnetic resonance (EPR) g-tensors;
\item[--] transport in nano-wires in the ballistic regime;
\item[--] near-edge X-ray absorption spectra;
\item[--] implicit solvation and electrochemical models;
\item[--] optical and electron energy loss spectra using time-dependent density-functional perturbation theory.
\end{itemize}

On the purely computational side, optimization for modern  and massively parallel high-performance computing (HPC) architectures has been pursued, using multiple parallelization levels involving both message-passing interface (MPI) and multi-threading (OpenMP).

All these advances, and related references, are documented in Refs.~\onlinecite{QE1} and \onlinecite{QE2}. In particular, Ref.~\onlinecite{QE2} also documents a preliminary version of package \texttt{HP}, performing linear-response calculations of Hubbard parameters, that has been since released.\cite{Timrov:2018}

In addition to being directly used for implemented property calculations, \qe\ is used in more creative ways as well:
\begin{itemize}[noitemsep,nolistsep]
\item[--] as a ``quantum engine'' for more sophisticated calculations,
  such as advanced MD calculations implemented in
  \texttt{i-Pi},\cite{KAPIL2019214} genetic and evolutionary algorithms,
  high-throughput calculations with \texttt{AiiDA};\cite{Pizzi:2016}
\item[--] in modified or patched versions, for specific
purposes, as in ``embedded'' \texttt{eQE}.\cite{doi:10.1002/qua.25401}
\end{itemize}
Finally, \qe\ is used to produce DFT data for further processing by other codes. We mention in particular codes performing Quantum Monte Carlo calculations
(\texttt{QMCPack})\cite{Kim_2018} and many-body perturbation theory (MBPT)
(\texttt{Yambo},\cite{Sangalli_2019}
\texttt{BerkeleyGW},\cite{DESLIPPE20121269}
\texttt{WEST},\cite{Govoni2015}
\texttt{Sternheimer-GW}\cite{Schlipf2020});%\cite{Lambert2013});
determining maximally localized Wannier functions 
(\texttt{Wannier90}\cite{Pizzi_2020}%\cite{Mostofi:2014}),
and electron-phonon interactions (\texttt{EPW}\cite{Ponce2016});
and codes performing different types of data analysis (\emph{e.g.} topological
analysis with \texttt{critic2})\cite{OTERODELAROZA20141007}
or computing transport properties (e.g.,
\texttt{WanT},\cite{Ferretti:2007}
\texttt{KGEC}\cite{CALDERIN2017118}).

Stable versions of the source code---the latest is v.6.5 at the time this article was written---as well as development and experimental versions can be downloaded from the URL: \url{http://www.quantum-espresso.org/download}.

\section{Challenges and opportunities of new heterogeneous architectures} 
\label{sec:towards-the-exascale}

The push of quantum mechanical materials simulation software towards ever increasing levels of complexity, accuracy, and performance has been so far assisted by a constant downsizing of micro-processing units, allowing for a steady increase of the compute capacity of general-purpose architectures at constant power. This process is now hitting its ultimate physical limits and is just about to come to an end. To reverse this state of affairs, major and disruptive changes in hardware architecture are to occur. The constraints set by power consumption can only be met by heterogeneous architectures, with specialized cores (``accelerators'') that maximize efficiency for a small set of instructions: \emph{e.g.}, graphics processing units (GPUs), PEZY chips, tensor processors, neuromorphic chips, \emph{etc.}. 
A quick look at the first places of the Top500 supercomputers list clearly shows that heterogeneous architectures have become the \emph{de facto} standard for new generation HPC systems.\cite{top500, feldman2018new}
On the single node scale, the ratio between the computational power provided by accelerators and traditional CPU units is found to grow at each  procurement cycle over the last 10 years.\cite{feldman2018new} 
Most of the computational power of future ``exascale'' machines (that is: capable of 1 exaflop, or $10^{18}$ floating-point operations per second) will come from accelerators.

The ensuing architectural complexity will set demanding requirements in terms of data movement, heterogeneous memory management, fault tolerance, which will all require a major, possibly joint, re-design of circuits and algorithms and the adoption of different programming paradigms.
In particular, extremely parallel applications will require rapid and substantial architectural shifts,  including, for example, the handling of intra-node data movement between disjoint memory spaces and the explicit treatment of deeper memory hierarchies.

Porting community codes to novel hardware architectures has always required extensive re-coding in the past. This cannot be sustained any longer in view of the considerable complexity reached by \qe\ and similar community codes (several hundred thousands code lines each) and the forthcoming diversity, heterogeneity, and rapid evolution of the hardware architectures. The solution we have identified is to refactor \qe\ into multiple layers, resulting from the assembly of weakly coupled components (modules and libraries), to be maintained and enhanced independently from each other, shared among different codes, and designed to be as architecture-agnostic as possible. A bird's eye view of the code will reveal four main such layers (see Fig.~\ref{fig:QEscheme}):
\begin{figure}
\includegraphics[width=0.99\columnwidth]{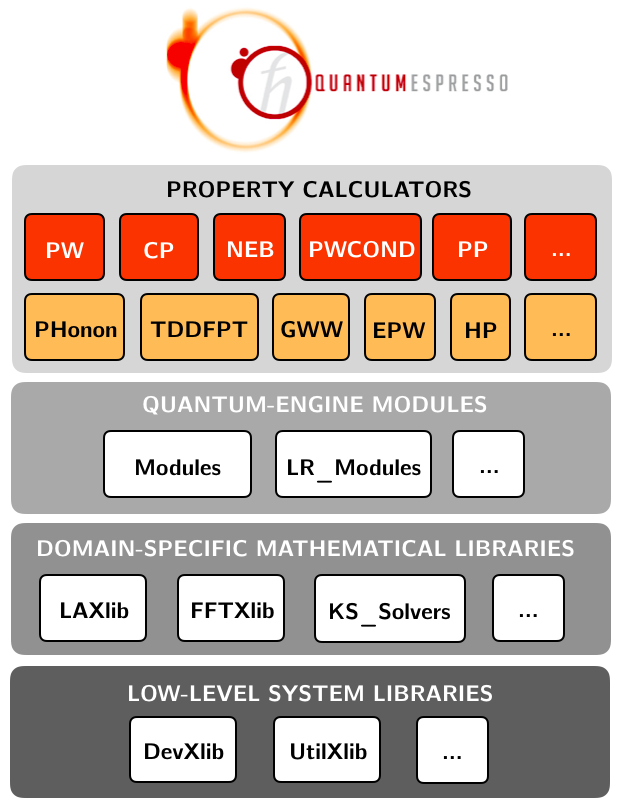}
\caption{Schematic illustration of the structure of the \qe \, distribution. The non-perturbative codes for SCF, MD, postprocessing and other calculations are those
in the first line of the ``property calculators'' section (highlighted in red), while the linear-response and MBPT codes are in the second line (highlighted in yellow). Low-level layers of the \qe \, distribution (as discussed in the text) are also shown.}
\label{fig:QEscheme}
\end{figure}
\begin{itemize}[noitemsep,nolistsep]
\item[--] The top layer is a collection of \emph{property calculators} to compute materials properties and to simulate processes, which is the ultimate goal of molecular and materials simulations. These property calculators may share global variables among themselves and with the quantum-engine modules (see below). This layer could be partially and progressively made code-agnostic, while it should remain as architecture-agnostic as possible.
\item[--] The core layer contains \emph{quantum-engine modules} that solve the one-particle Schr\"odinger equation and perform associated tasks, such as Hamiltonian builds (\emph{i.e.} the application of the one-particle Hamiltonian to molecular orbitals and Bloch states), or other tasks related to density-functional, linear-response, and many-body perturbation theories. This layer is and will likely continue to be code-specific, but should stay as architecture-agnostic as possible. Ideally, the same quantum-engine modules could be shared among different codes performing similar tasks, \emph{e.g.} standard SCF computations and geometry optimizations, \emph{ab initio} MD, and linear-response/MBPT calculations.
\item[--] A collection of \emph{domain-specific mathematical libraries}, to perform general-purpose tasks, such as 3D Fourier analysis, linear algebra, using both iterative and factorization algorithms, non-linear optimization, \emph{etc.} Such mathematical libraries should be easily shared among different codes of a same class (\emph{e.g.} adopting the same quantum-mechanical representation / basis set) and stay largely architecture-agnostic as above. While these libraries may themselves contain modules, in order to ensure easy portability across different codes, the data exchange between them and the calling code will only occur via standard public application programming interfaces (APIs), which will make minimal use of complex data structures and derived data types.
\item[--] Finally, a collection of \emph{low-level system libraries}, which will abstract the most recurrent architecture-specific constructs, such as data offload to/from specialized hardware, memory synchronization, and the like. These libraries will interface directly with the underlying hardware architectures and should be fine-tuned on each of them for optimal performance. They should be architecture-specific, and as much code-agnostic as possible, without interfering with the other layers.
\end{itemize}

This strategy is meant to accomplish what has come to be dubbed \emph{separation of concerns}: ideally, method developers in science departments and research labs should be concerned with property calculators, disregarding architectural details, whereas scientists and research engineers in IT departments and HPC centers should focus on low-level mathematical and system libraries. The two categories of developers should work shoulder-to-shoulder to develop and maintain the quantum engine. Separation of concerns is the overarching guideline for the action of the EU \MaX\ Centre of Excellence (CoE) for HPC Applications,\cite{MaX} whose mission is to foster the porting of important community codes for quantum materials modeling to heterogeneous architectures.

The implementation of the above strategy is bound to entail an extensive refactoring of community codes, which in turn will require a longer time than allowed by the pressing advance of heterogeneous architectures. For this reason, the course of action undertaken by the \qe\ developers follows two converging tracks. On the one hand, an accelerated version of \texttt{pw.x} working on NVIDIA GPUs is already available, and more components---initially \texttt{cp.x}, followed by \texttt{ph.x} and other linear-response codes---are being ported. On the other hand, the refactoring of the whole distribution into a collection of stand-alone modules and libraries is steadily progressing. 
The final goal to which both tracks will eventually converge is the identification and gradual encapsulation of architecture-specific computational kernels into low-level libraries, so as to make the migration to multiple architectures as seamless as possible.
We aim in particular at using the forthcoming OpenMP-5 standard, that in addition to architecture-agnostic APIs (already in OpenMP 4.5) provides deep memory copies and a simple framework for exploiting unified shared memory when available.

In the following we discuss in some detail the different logical layers that constitute the structure of the \qe \, distribution, as shown in Fig.~\ref{fig:QEscheme}, starting from the bottom layer. 
Low-level system libraries are discussed in Sec.~\ref{subsec:performance_portability} while the remaining layers
are discussed in Sec.~\ref{subsec:SustainableSoftwareDevelopment}.

\subsection{Performance portability}
\label{subsec:performance_portability} 
Performance portability across current and future heterogeneous architectures is one of the grand challenges in the design of HPC applications. General-purpose frameworks have been proposed,\cite{kokkos, alpaka2016, alpaka2017, raja} but none of them has reached maturity and widespread adoption. In addition, Fortran support is still very limited or missing entirely. In this context, the \MaX\ CoE is promoting and coordinating a collective effort involving the developers of various materials modeling applications.
Taking on this challenge with a domain-specific approach has the advantage of providing abstraction and encapsulation of a limited number of functionalities that constitute the building blocks of the most common operations performed on the accelerators in this field. This will allow us to prepare low-level architecture-specific implementations of a limited number of kernels that have been already characterized and isolated, thus keeping the source code of the various scientific applications untouched and reducing code branches when new systems will appear on the market. 

Such an effort is still in the early stages, but is under active development and is progressively entering the GPU port of \qe{} through the so-called \texttt{DevXlib} library.
This library started off as a common initiative shared among \MaX\ CoE codes (notably \qe{} and \texttt{Yambo}), aimed at hiding CUDA Fortran extensions (see Sec.\ref{subsec:GPUVersion}) in the main source base. Being used by different codes, the library has been rationalized and further abstracted, thus becoming a performance portability tool aimed at supporting multiple back-ends (support to OpenACC and OpenMP-5 foreseen, direct extension to CUDA C possible).
The main features included in the library by design are the following:
\begin{itemize}[noitemsep,nolistsep]
  \item[--] performance portability for Fortran codes;
  \item[--] deal with multiple hardware and software stacks, programming models and missing standards;
  \item[--] wrap/encapsulate device specific code;
  \item[--] focused on limiting code disruption (to foster community support).
\end{itemize}
It is important to note that part of the library design includes the definition of which device-related abstract concepts need to be exposed to the scientific developers. To give an example, memory copy and synchronization to/from host/device memory are abstract operations that the developers of property calculators or of the quantum engine itself may need to control directly. Therefore, \texttt{DevXlib} exposes such control in the form of library APIs that are agnostic of the specific device back-end.

In practice, \texttt{DevXlib} provides the user with ($i$) interfaces to {\it memory handling} operations including creation and locking of memory buffers  (\texttt{device\_memcpy} and \texttt{device\_buffers}); ($ii$) interfaces to {\it basic and dense-matrix linear algebra routines}, similarly to BLAS and Lapack (\texttt{device\_linalg}); ($iii$) interfaces to more {\it domain-specific operations} (\texttt{device\_auxfuncs}); ($iv$) device-oriented {\it data structure} compatible with Fortran usage.
In particular, memory handling allows the user to copy memory host-to-host, device-to-device, and also across memories, host-to-device and vice-versa, thereby dealing also with memory off-load and synchronization. Importantly, both synchronous and asynchronous copies can be performed with explicit control. 
Moreover, the explicit handling of memory buffers is meant to ease or avoid the procedure of allocation and deallocation of auxiliary workspace. 

Besides the interface to linear algebra optimized libraries, such as cuBLAS and cuSOLVER in the case of NVIDIA GPUs, \texttt{DevXlib} also provides interfaces to selected (more specialized) kernels that appear often in plane waves electronic structure codes (such as scalar products with $\mathbf{G}$-vector remapping or specific matrix or vector updates, to name a few). While not covering all possible kernels of the same complexity, this is quite effective in helping to keep the code sources as clean as possible.
Importantly, some efforts have also been directed to devise Fortran-compatible data structures to handle the extra complexity of data stored on host and/or accelerator memories. In principle, these data structures need to be able to hide hardware complexity (e.g. being vendor agnostic for what concerns accelerator memory), to allow for seamless memory synchronization and transfer, and to be equipped with a set of methods to implement the most common data operations. Software engineering and experimentation of these data structures is currently ongoing. 

In order to make the code more flexible and portable, some basic functionalities are accessed via common interfaces provided by a low-level library \texttt{UtilXlib}, including utilities for MPI and OpenMP parallelization, timing, error and memory handling. This library has been extended to include execution and data synchronization points for communications involving memory spaces located on the GPUs.

\subsection{Towards a sustainable development, maintenance, and porting model} \label{subsec:SustainableSoftwareDevelopment}

\qe{} has grown in size during the years, including as of v.6.5 almost 600,000 lines of Fortran code, 60,000 lines of C, python, shell scripts, plus a large amount of tests, examples, and documentation. While not especially large with respect to scientific software projects in other fields, \qe{} is sufficiently bulky to make its maintenance and extension a challenge that cannot be sustained without resorting to modern software engineering techniques. Much work, described in part in Sec.~3 of Ref.~\onlinecite{QE2}, has been done in the past along the following directions:
\begin{itemize}[noitemsep,nolistsep]
\item[--] provide a large set of automated tests to ensure the validity and portability of the results under different building and execution patterns;
\item[--] extend the interoperability with other codes via structured I/O, using an extended markup language (XML) with an industry-grade \emph{schema} description for small human-readable, data files, and optionally a hierarchical format (HDF5) for large, binary, data sets;
\item[--] collect large parts of the  code base into modules and libraries, in order 
to enhance its readability, ease of maintenance, and portability.
\end{itemize}
%{\bf Definition of libraries and modules.} 
The work along the latter direction has been extended with the creation of distinct code layers: ($i$) \emph{Libraries} and ($ii$) \emph{Modules}. {\it Libraries} have a well-encapsulated inner data structure and exchange data with the rest of the code only through predefined APIs. As a consequence, libraries can be developed, distributed, and compiled independently of each other and then linked to the main code. \emph{Modules} are reusable blocks of code whose functionalities are accessed via well-defined APIs as well but, mainly for reasons of computational efficiency, do not stick to such a clear-cut data encapsulation scheme as libraries and share a significant portion of their data structure with the rest of the code. For this reason, modules must be compiled together with the main code and are mainly intended for usage inside the \qe\ suite or other codes sharing the same global data structure.

\subsubsection{Domain-specific mathematical libraries}
\label{sec:domain_specific_libs}

Apart from improving the  maintainability of the whole distribution, packaging the code base into libraries also has the advantage of making distributed development and maintenance easier and of providing a wider community with effective tools for developing electronic-structure software ready to use in modern HPC infrastructures.
% and the possibility to be developed and maintained by a wider community.  
For this reason we aim at avoiding the usage of structured data type as arguments of the APIs as much as possible and at exposing the  interfaces using  included files rather than Fortran modules. 

Currently, three major packages have been extracted from \qe{} and are ready to be distributed as stand-alone libraries, namely:
\begin{itemize}[noitemsep,nolistsep]
\item[--] \texttt{LAXlib}, performing parallel dense-matrix operations, including basic linear algebra and diagonalization; %, matrix-matrix, and matrix-vector multiplications;
\item[--] \texttt{FFTXlib}, for parallel distributed three-dimensional fast Fourier transforms;
\item[--] \texttt{KS\_Solvers}, a collection of iterative diagonalization algorithms to solve the Kohn-Sham equations.
\end{itemize}
A large part of the computations of a typical electronic-structure calculation is performed inside these libraries. The usage of machine-optimized mathematical libraries and the inclusion of further optimizations, either architecture-agnostic or architecture-specific, in these libraries will automatically profit to all codes and computations. It is at this level that the separation of concerns is most fruitful in terms of performance portability. While the original code targeted MPI and OpenMP parallelization on many CPUs, the extension to different programming paradigms for heterogeneous architectures has much progressed since, also thanks to contributions from IT experts. 

\texttt{LAXlib} and \texttt{FFTXlib} libraries, with their completely encapsulated  inner data structures, can be easily used by third parties. Their interfaces  are  transparent to the specific underlying architecture. 

The iterative diagonalization algorithms collected inside the \texttt{KS\_Solvers} are disentangled from the specific Hamiltonian builder, which is called by the library as an external routine; the definition of wavefunctions and their scalar products inside the  Hamiltonian builder must be compatible with the one used inside \texttt{KS\_Solvers}. For some of the algorithms, a Reverse Communication Interface (RCI) is also available, allowing one to directly pass the $H\left | \psi\right.\rangle$ vectors to the library, leaving to the programmer the task of computing and converting them to the format expected by the RCI. 

The goals of the activities described here largely overlap with those of the Electronic Structure Library (ESL) project at CECAM\cite{ESL}. The possibility of decoupling the \texttt{KS\_Solvers}, \texttt{LAXlib} and \texttt{FFTXlib} libraries from their native codes was first demonstrated during a Workshop organized in 2017 within the ESL initiative\cite{ESLW_Drivers}. Moreover, both \texttt{KS\_Solvers} and \texttt{LAXlib} may use another library maintained by ESL, \texttt{ELPA} (included in \texttt{ELSI}) for dense-matrix diagonalization. \qe{} may also use the \texttt{libxc} ESL library computing exchange-correlation functionals.

\subsubsection{Quantum-engine modules}

For other parts of the code, data encapsulation is difficult to achieve or even unfeasible, or may introduce inefficiencies. For those cases, it was chosen to refactor the code into general modules, still using the global data structure of the suite. These modules are currently meant to be used inside the distribution, but they are designed to be easily accessible for the development of further applications built upon the \qe{} code base. Notable examples are \texttt{Modules} and \texttt{LR\_Modules}.

\texttt{Modules} is a collection of Fortran modules and subroutines that implement various operations needed to solve self-consistently and iteratively the Kohn-Sham equations of DFT. In particular, \texttt{Modules} contains the following functionalities: $(i)$~definition of global variables and universal constants, 
$(ii)$~reading of input parameters and of pseudopotentials, 
$(iii)$~definitions of Bravais and reciprocal lattices,
$(iv)$~symmetry analysis and symmetrization operations, 
$(v)$~calculation of the exchange-correlation potential, 
$(vi)$~generation of plane waves and of their mapping to FFT grids,  $(vii)$~generation of $\mathbf{k}$ points, 
$(viii)$~calculation of pseudopotential terms. Historically, \texttt{Modules} exists since the very first version of \qe{}, but it has continuously evolved in order to adapt to novel utilities and packages of the suite.

\texttt{LR\_Modules} is a much more recent part of \qe{}, which appeared about five years ago and evolved significantly since that time. The reason for the creation of \texttt{LR\_Modules} was to unify, harmonize, generalize, and refactor the functionalities that are common to all linear-response and MBPT codes of the suite. \texttt{LR\_Modules} contains the following functionalities:
$(i)$~definition of global data structures for linear response, 
$(ii)$~calculators of linear-response quantities (such as e.g. response density and potentials), $(ii)$~iterative solvers (e.g. Lanczos recursive algorithms), 
$(iii)$~response exchange-correlation kernel calculators, 
$(iv)$~symmetrization routines, 
$(v)$~projectors on the empty-states manifold, to name a few. 
The functionalities of \texttt{LR\_Modules} are used in the following packages: 
\begin{itemize}[noitemsep,nolistsep]
\item[--] \texttt{PHonon} for calculation of lattice vibrational modes (phonons), Born effective charges, dielectric tensor, 
and other vibrational properties;~\cite{BGT, Giannozzi:1991, deGironcoli:1995, DalCorso:1997, Baroni:2001, DalCorso:2000, DalCorso:2001, DalCorso:2007, DalCorso:2010, Floris:2011, Sabatini:2016, Sohier:2017, Urru:2019, Floris:2020}
\item[--] \texttt{TDDFPT} for calculation of optical absorption spectra of molecular systems,~\cite{Baroni:2012, Walker:2006, Rocca:2008, Malcoglu:2011, Ge:2014, Timrov:2015b} collective excitations in solids such as plasmons~\cite{Timrov:2013, Timrov:2015, Motornyi:2020} and magnons;~\cite{Gorni2018} 
\item[--] \texttt{EPW} for calculation of electron-phonon coupling, transport, and superconducting properties of materials;~\cite{Ponce2016} 
\item[--] \texttt{HP} for the first-principles calculation of Hubbard parameters of the Hubbard-corrected DFT.~\cite{Timrov:2018} 
\end{itemize}
The generalized and unified subroutines from \texttt{LR\_Modules} have been refactored in such a way that they can be easily and straightforwardly employed in any other future linear-response or MBPT code of \qe \, or even in third-party codes. They can now be used generically to build perturbations, apply them to the occupied ground-state Kohn-Sham wave functions and compute the related self-consistent first-order response properties either by solving the Sternheimer equations or by solving the Liouville quantum equations using the Lanczos recursion method.

\subsubsection{Interoperability}

Exportable output in \qe{} is based on the adoption of standard data formats: XML and HDF5. These two formats have the advantage of providing information about the hierarchy and the types of the data that  may be automatically processed by external applications. The support of these features in modern scripting languages like python makes them  convenient for the development of  postprocessing and  analysis tools.
For HDF5 the description of the hierarchy and of data types is contained in the file; for XML files we provide it under the form of XSD schemas.~\cite{xsd} In order to streamline the reading of XML files in python using the specifications of the schemas we have also released a python package, \texttt{xmlschema},~\cite{xmlschema} that  converts the ASCII content of the XML file into a corresponding python dictionary whose structure and data types follow  the indications given by the schema. The python dictionary  can then be used directly by the reading application or saved as JSON of YAML files. 

The coherence between the released  schemas for XML and the effective output of the applications is guaranteed by a python set of tools that produce Fortran bindings for reading and writing XML data, starting from the XML schema.  The tools generate the writing routines from the format specification and keep them automatically aligned with the schema. These tools have also been released as a separate package.\cite{xmltool}

The  \texttt{qeschema}~\cite{qeschema} package provides the APIs  specific for reading and writing the XML, HDF5 and unified pseudopotential format (\texttt{UPF})  files used by  the applications in the suite. This package also provides converters to map some structured data as \emph{e.g.} crystal structures to and from other format commonly used in visualizers, structure databases or atomistic simulation utilities as \texttt{ASE}~\cite{ase} or \texttt{pymatgen}~\cite{pmgen} or HDF5 charge densities which  may be exported to other formats, for instance the XSF format of \texttt{xcrysden}.~\cite{xcrysden} 

The \texttt{postQE}~\cite{postqe} package provides  python extensions and APIs to use the postprocessing utilities of the \texttt{PP} package inside python applications. Many \texttt{PP} components are   compiled as python extensions using \texttt{F2PY}.  For those applications that may have a large computational load and  memory footprint, 
\texttt{postQE} provides instead  tools to extract data from the output files of the Fortran executables. 
 
\subsection{Evolution of the GPU-enabled version}
 \label{subsec:GPUVersion}
 
\begin{figure*}
\includegraphics[width=1.95\columnwidth]{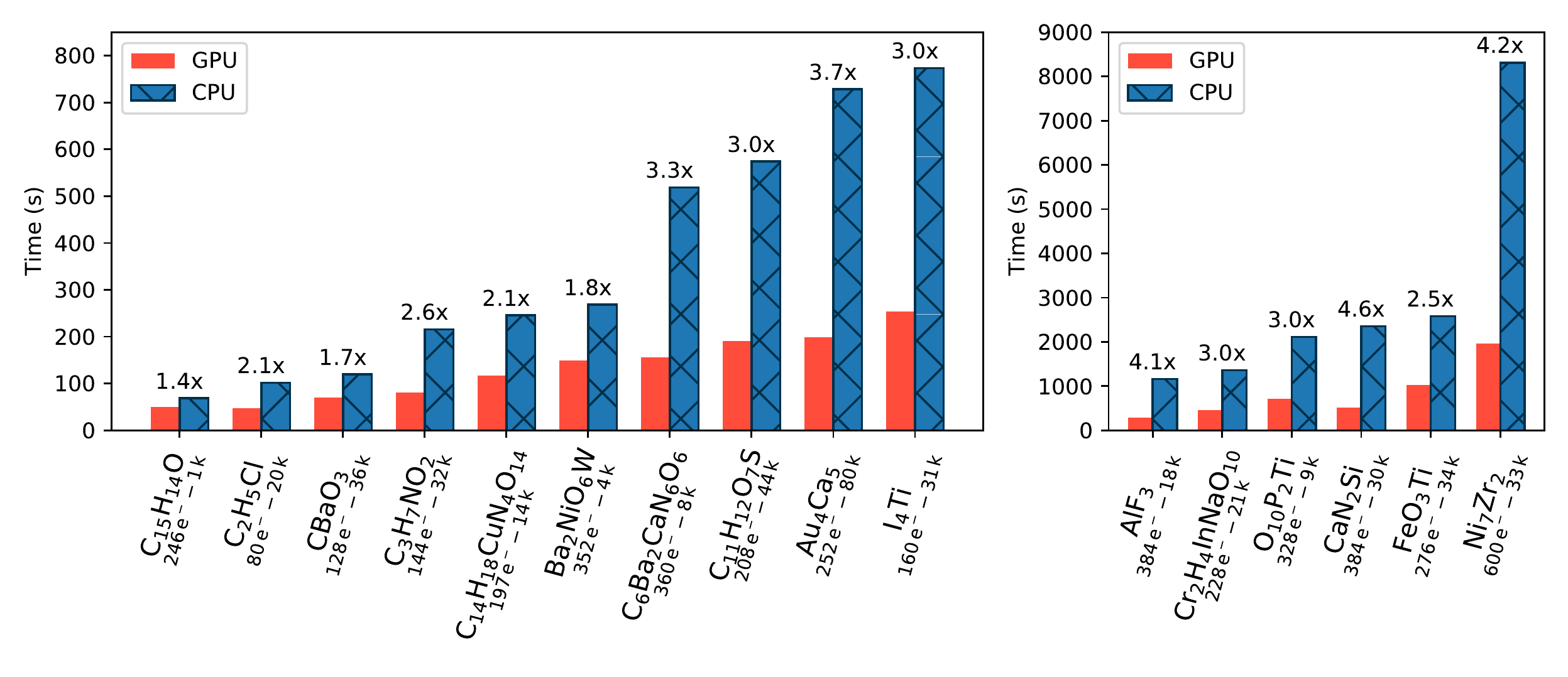}
\caption{Collection of benchmarks performed with \qe{} v.~6.5 and QE-GPU v6.5 on a
18-core Intel(R) Xeon(R) E5-2697 v4 @ 2.30GHz or the same CPU plus a NVIDIA V100 card.
The number of electrons per simulation cell and the number of $ \mathbf{k}$-points used for reciprocal space sampling is reported close to the chemical formula. For each simulation the sum of the time taken by the initialization, the iterations during self consistency and the estimation of atomic forces is compared.
\label{fig:gpuperf}}
\end{figure*}

\qe\ introduced support for accelerated systems as early as 2013, starting from v.~5.0, in the form of custom plugins for some of the codes of the suite.\cite{Spiga2012, qe-gpu-plugin} This initial approach, based on CUDA C and \emph{ad hoc} libraries for linear algebra,\cite{Spiga2012} proved to be successful in boosting the code performance,\cite{girotto2012enabling} but hardly
sustainable from the maintenance and development 
points of view, mainly due to the substantial amount of ``boilerplate''  (replicated) code used to interface Fortran subroutines with CUDA C.

In light of this limitation, a new port, dubbed QE-GPU, has been recently rewritten from scratch, starting from the case study presented by Romero \emph{et al.} \cite{romero2017performance} who ported v.~6.1 of \texttt{pw.x} to NVIDIA GPUs.\cite{GPUacceleratedQuantumESPRESSOZenod}
In Ref.~\onlinecite{romero2017performance}, the authors detail a new strategy based on CUDA Fortran---the Fortran analogue of CUDA C---and  demonstrate $2\times$ to 3$\times$ speedup consistently achieved on a variety of platforms and using different benchmarks. The new GPU-enabled releases of \qe\ extend this work, but adopting a few design solutions to streamline future development and porting to other heterogeneous architectures, as detailed below.

While still being developed with CUDA Fortran, the last release of QE-GPU is almost entirely accelerated through a directive-based approach, using the so-called \emph{cuf kernel} compiler directive, that generates parallel architecture-specific code for loop-based structures. In some cases this choice may limit the code performance, but it brings a number of positive consequences. Firstly, it allows one to validate GPU code on the CPU and, in principle, to retain a single source code for both CPU and GPU implementations. Second, this design choice enables the straightforward adoption of other directive-based programming models, like OpenACC or OpenMP, if required.
%OpenACC was found to be a valid option, but explicit memory management and
%performance was the reason behind our first choice in this case.
%OpenMP is potentially the only alternative for creating
%portable accelerated code, but the current lack of feature complete and robust
%implementations of the 4.5 and 5.0 standards limits its adoption.
As a consequence, even if the current implementation only runs on NVIDIA GPUs, special care has been paid to design the software in a way that minimizes the future effort needed to port the code to other accelerators that can be exploited through a directive-based programming model. In particular, attention has been paid not to introduce CUDA Fortran kernels in the high-level structure of the code, except for trivial data transformations, and to keep them confined to 
the domain-specific and system libraries described in Secs.~\ref{subsec:performance_portability} and \ref{sec:domain_specific_libs}.

A further advantage of CUDA Fortran is explicit memory management. \qe{} organizes related data into modules and derived types. Allocation and synchronization of these kinds of data structures is a straightforward and concise operation in CUDA Fortran, while OpenACC and OpenMP support automatic synchronization of these data types only in the most recent releases. In addition, by retaining full control on memory allocations and synchronizations between the RAM and GPU's global memory, we provide a clear view of all intra-node memory operations to the programmer, thus facilitating future porting activities.

As detailed in Sec.~\ref{subsec:SustainableSoftwareDevelopment}, a few specific components of \qe{} have been extracted from the main code trunk, made independent from it and released as packaged libraries. Some of them have been targeted for GPU acceleration, namely the parallel dense eigenvalue solver (\texttt{LAXlib}), and the parallel distributed FFT (\texttt{FFTXlib}). In this case we abandoned directive based acceleration in favor of architecture specific APIs and libraries. Indeed, the last release of these libraries relies on cuBLAS for linear algebra kernels, cuFFT for 1D and 2D Fourier transforms, and cuSOLVER for solving real and complex generalized eigenvalue problems.\footnote{Optionally, cuRAND can also be used for the generation of random wave-functions.}

As of version 6.5, the following set of functionalities benefit from GPU acceleration and operate on data residing on the GPU:
\begin{itemize}[noitemsep,nolistsep]
\item[--] electronic self-consistency for both magnetic and spin-unpolarized systems;
\item[--] iterative solution of the Kohn-Sham Hamiltonian using either the Conjugate Gradient or the Davidson method;
 \item[--] calculation of atomic forces;
 \item[--] calculation of exact exchange terms for hybrid functionals.
\end{itemize}
The acceleration is obtained by exploiting numerous kernels that have been ported to GPU: local-potential, pseudopotential, kinetic-energy contributions to the Hamiltonian, 
preconditioner evaluation for iterative solvers,
application of the overlap operator, 
wavefunction initialization,
Hubbard component contribution to the effective potential,
charge density generation.
For all the aforementioned operations, where applicable, both the case of real-valued wavefunctions
(${\bf k}=0$ only sampling, useful to reduce the memory footprint and to speed up simulations)
and the case with spinor wavefunctions (non-collinear magnetism) have been ported to GPU.

Currently, only \texttt{pw.x} can benefit from GPU acceleration,
but other codes are being ported and will be available in future releases.
The GPU-enabled version of \texttt{pw.x} is fully compatible with its CPU counterpart, provides the
same features, undergoes the same regression testing suite, and converges to equivalent results within a given convergence criterion.
%As a consequence, all postprocessing tools can already be used to analyze the
%results of GPU accelerated simulations and restart files work smoothly
%across the two versions.
%

%
The speedup provided by the GPU implementation depends drastically both on the hardware and on the details of the input data. The extreme scale performance has been already detailed elsewhere,\cite{maxdeliv_D4.2,maxdeliv_D2.1}  thus here we focus on smaller problems, consisting of tens of atoms and hundreds of electrons. This will allow us to identify the lower limit for the input cases that can benefit from GPU acceleration. 
% thus providing important information for the high-throughput like workloads.
%Second, it probably resembles more closely the most common everyday practices of QE users.

The \texttt{qeinputgenerator}~\cite{qeinpgen}  was used to prepare a set of benchmarks. This tool automatically generates input files for \texttt{pw.x} providing a few options to customize the accuracy of the simulation and using pseudopotentials from two sets,
either standard solid-state pseudopotentials (SSSP) Efficiency or SSSP Precision.\cite{Prandini2018} These in turn include norm conserving,\cite{Willand2013,Schlipf2015,vanSetten2018} ultrasoft,\cite{Garrity2014} and PAW \cite{DalCorso2014,Topsakal2014} pseudopotentials thus covering a significant fraction of \texttt{pw.x} functionalities. Sixteen structures having unit cell volume between 250~{\AA$^{3}$} and 1000~{\AA$^{3}$} were randomly selected from the Crystallography Open Database (COD).\cite{Quiros2018, Merkys2016,Grazulis2015, Grazulis2012,Grazulis2009,Downs2003} Structures with fractional occupations and rare earth elements were discarded.
All input and output data are 
available on the Materials Cloud archive.~\cite{MaterialsCloudArchive2020} 
% PB: I moved technical details about the inputs and the results here: https://gitlab.com/bonfus/qe-gpu-benchmarks/-/blob/master/README.pdf}

%THIS GOES TO SI: according to the following syntax: \textsc{ID-PseudoSet-OccupationType-ReciprocalSpaceGrid}. \textsc{ID} is COD's structure ID, \textsc{PseudoSet} can either be ``E'' or ``P'' for SSSP Efficiency or Precision respectively. This choice obviously implies different values for the cutoff of the basis set. The \textsc{OccupationType} can be NI for ``non-magnetic insulator'' having fixed electronic band occupations, NM for ``non-magnetic metal'' where a smearing function \cite{PhysRevLett.82.3296} is used for bands occupation close to the Fermi level or ``MM'' for system where spin-polarization was considered. The \textsc{ReciprocalSpaceGrid} parameter can be ``N'' for grids with 0.3 \AA$^{-1}$ distance between reciprocal space points, ``F'' for 0.2 \AA$^{-1}$ and ``V'' for 0.15 \AA$^{-1}$. In addition, in a few cases, we have manually enabled the so-called gamma-trick and set this label to ``G''.

In Figure~\ref{fig:gpuperf} we compare the best time-to-solution obtained with a single 18-core Intel(R) Xeon(R) E5-2697 v4 @ 2.30GHz (BDW) CPU and the same hardware accompanied by one NVIDIA's Tesla V100 GPU card. The ratio between the theoretical peak performance of these two units is roughly 1 to 10, but effective GPU acceleration can only be achieved with extremely data parallel workloads and the speedup provided by the graphic card can even become lower than 1 when this condition is not met. This limit is investigated with the present benchmark using a rather standard balance between GPU and CPU computational power for a HPC node.

The CPU results have been collected with \qe \ v.~6.5 compiled with Intel's 2019 suite, Intel's MPI implementation and Intel's Math Kernel Library (MKL), while QE-GPU v.~6.5 was built with the PGI 19.10 Fortran compiler and linked to Intel's 2019 MKL. For the set of inputs detailed above, the pure MPI parallelization is the best strategy for the CPU version of the code, therefore we performed all simulations for the CPU version disabling OpenMP. On the other hand, the GPU code requires OpenMP parallelism since the number of MPI processes is limited by the number of GPU cards installed on the system: each MPI process should be assigned just one accelerator.\footnote{Although as a rule the code should be executed with one MPI per GPU card, over-subscription of the GPU can lead to significant speedups, especially for small test cases. Nonetheless, over-subscription by more than a factor 4 is seldom useful.} For both the CPU and the GPU versions, only parallelism over $\mathbf{k}$-points (\textsc{-npool} option) has been used: a reasonable choice, given the relatively small dimension of the dense eigenvalue problem to be solved during iterative diagonalization.

The two sets of results are numerically equivalent: the largest discrepancy between the total energies computed by the CPU and the GPU versions is $2\cdot 10^{-8}$ Ry, while the largest difference in the total force is $3\cdot 10^{-5}$~Ry/Bohr.

In Fig.~\ref{fig:gpuperf} we report the wall time required to reach convergence (or 80 SCF cycles when convergence was not achieved) for the CPU and GPU versions. Only the best time to solution as a function of $\mathbf{k}$-point parallelism is reported. The total execution time has also contributions, not reported because much smaller, from the initialization step and by the final calculation of atomic forces.
%These three timers describe the whole execution for the tests that we selected.
%
For the smallest test case, a 98-atom organic compound with 246 electrons that converges in less than 1 minute on our hardware, the GPU speedup is just 1.4x.
This is indeed the current limiting size for taking advantage of GPU acceleration. On the other hand, as the problem size grows, especially as a function of the basis set dimension,\footnote{This can be appreciated by comparing simulations that required similar computational time on the CPU but used different pseudopotential sets, for example the O$_{10}$P$_2$Ti and CaN$_2$Si columns on the right panel of Fig.~\ref{fig:gpuperf}.} the speedup can exceed 3x.
\section{Outlook and conclusions}
\label{sec:Conclusions}

The work on the \qe\ distribution has since many years followed some well-established lines of action: $(i)$ implementing new numerical algorithms, theoretical methods, property calculations; $(ii)$ improving  interoperability with external codes; $(iii)$ making the codes more modular and easier to work with, more portable across architectures, without sacrificing performances. The arrival of heterogeneous, accelerated architectures has made the latter direction more urgent and wider in scope. It is no longer sufficient to isolate low-level computational kernels into simple mathematical libraries: performance portability must be ensured at the level of so-called ``domain-specific'' libraries. Without effective actions in the direction $(iii)$, it will become difficult to implement, and even more difficult to maintain, further property calculations, new theories, and new algorithms.

The work described in Sec~\ref{subsec:GPUVersion} is the starting point for the effective implementation 
of the plan described in Sec.~\ref{sec:towards-the-exascale}. The next significant step is the merger of 
the main \qe{} distribution with the CUDA Fortran version for NVIDIA GPUs. This will eliminate the constraint
of keeping the latter version akin to the main distribution in order to simplify the ``alignment'' process as 
the main distribution evolves.

In parallel, more work is ongoing to achieve performance portability, as described in
Sec.~\ref{subsec:performance_portability}. In particular, support for OpenMP-5 
is being introduced into the domain-specific mathematical libraries of
Sec.~\ref{subsec:SustainableSoftwareDevelopment} and the low-level system libraries of
Sec.~\ref{subsec:performance_portability}. 
In this respect we are working with Intel software engineers to guarantee complete compatibility for their future Ponte Vecchio GPU architecture, and \qe{} will be ready to run on these cards when they will appear on the market, towards the end of 2021\cite{Aurora}. Porting to AMD accelerated architectures is also ongoing exploiting the public OpenACC stack. Work on ARM vectorized (SVE instruction set) architectures 
(EPI\cite{EPI}  and AFX64\cite{Fugaku} processors) is also
ongoing and on track to release an optimized \qe{} version for these architectures. All these porting efforts would be hardly feasible without the
contribution of IT experts and HPC centers, in the spirit of the separation of concerns. This solution does not target the maximum level of architecture-specific optimizations,
but leaves the possibility open to achieve them, once the hardware is finally installed in major HPC centers,
using specially-tailored patched versions. 

The benchmarks presented in Sec.\ref{subsec:GPUVersion} are realistic but relatively small-size use cases. 
The extension to large-scale calculations requires further work to identify and remove memory and computation bottlenecks. In particular, the amount of global memory per card is a relevant parameter that substantially
impacts the performance of the accelerated version of the code. Some of the largest benchmarks in our set do saturate the memory available on a single card, depending upon the parallel options used to distribute 
plane-wave components. Similar issues will certainly show up also in extreme scale benchmarks.
For single-point self-consistent or molecular-dynamics calculations, the most critical bottlenecks towards scaling to the whole exascale system will be the amount of memory 
in cards, due to the superlinear complexity of plane-wave electronic structure
calculations (the amount of memory and computation increases more than quadratically 
with the size of the system).
It is foreseen that the best ``exaflop'' performances will be actually achievable for 
selected cases, such as ``high-throughput'' and ``ensemble'' large-scale calculations, that can be split into many smaller ones. The efficiency of each computation will 
thus be critical for fully exploiting the capability of the HPC systems.

{\bf Data availability statement}.
The data that support the findings of this study are openly available in the Materials Cloud Archive at \url{https://doi.org/10.24435/materialscloud:2020.0021/v1}.\cite{MaterialsCloudArchive2020}

\begin{acknowledgments} This work was partially funded by the EU through the \MaX\ Centre of Excellence for HPC applications (Project No. 824143). Financial and administrative support from the \qe{} Foundation\cite{qef} is also gratefully acknowledged. N.M. and I.T. acknowledge support from the Swiss National Science Foundation (SNSF), through grant 200021-179138, and its National Centre of Competence in Research (NCCR) MARVEL. Valuable technical support from NVIDIA is gratefully acknowledged. \qe{} has received contributions from many people in addition to the authors of this article. In particular, we would like to thank the following colleagues who have contributed to different stages of the design and implementation of the QE-GPU code: 
Fabio Affinito, %(CINECA)
Anoop Chandran, 
Brandon Cook, % (NERSC),
Massimiliano Fatica, % (NVIDIA),
Ivan Girotto, % (ICTP),
Thorsten Kurth, % (NVIDIA)
Miloš Marić, % (NVIDIA),
Everett Philips, % (NVIDIA),
Josh Romero, % (NVIDIA),
	and Filippo Spiga. % (ARM),

\end{acknowledgments}

%\bibliographystyle{unsrt}
%\bibliography{main}

\end{document}